\documentclass[preprint2]{aastex}

\slugcomment{Accepted, $ApJlett$,
December 20, 1999}

\shorttitle{Rings in $\beta$ Pic}
\shortauthors{Kalas et al.}

\begin{document}


\title{Rings in the Planetesimal Disk of $\beta$ Pic}


\author{P. Kalas\altaffilmark{1}}
\affil{Space Telescope Science Institute}
\email{kalas@stsci.edu}

\author{J. Larwood\altaffilmark{2}}
\affil{Queen Mary \& Westfield College}
\and
\author{B. A. Smith\altaffilmark{3}}
\affil{University of Hawaii}
\and
\author{A. Schultz\altaffilmark{1}}
\affil{Space Telescope Science Institute}


\altaffiltext{1}{Space Telescope Science Institute, 3700 San Martin Dr., Baltimore,
MD  21218}
\altaffiltext{2}{Astronomy Unit, Queen Mary \& Westfield College, London E1 4NS, UK}
\altaffiltext{3}{Institute for Astronomy, University of Hawaii,
2680 Woodlawn Dr., Honolulu, HI  96822}


\begin{abstract}
The nearby main sequence star $\beta$ Pic is surrounded by an
edge-on disk of dust produced by the collisional erosion
of larger planetesimals.
Here we report the discovery
of substructure within the northeast extension of the disk midplane 
that may represent an asymmetric
ring system around $\beta$ Pic.
We present a dynamical model showing that a close stellar
flyby with a quiescient disk of planetesimals 
can create such rings, along with  
previously unexplained disk asymmetries. 
Thus we infer that $\beta$ Pic's planetesimal disk was
highly disrupted by a stellar encounter in the 
last hundred thousand years.
\end{abstract}




\section{INTRODUCTION}
Circumstellar dust disks around nearby stars provide strong
indirect evidence for the existence of planetesimals in exosolar
systems. Radiation pressure, Poynting-Robertson drag,
collisions, and sublimation remove dust orbiting young main sequence
stars on timescales 10$^2$ - 10$^3$ times shorter than the
stellar ages \citep{bap93}. Therefore we infer that larger parent bodies
exist and replenish dust in the same way that comets and asteroids
resupply interplanetary grains in the Solar System.  
 
We expect that an undisturbed system of many bodies orbiting a star will
have an axially symmetric distribution.  Over large (10$^2$ AU) scales, however,
the northeast (NE) side of the $\beta$ Pic disk
is not a mirror image
of the southwest (SW) side \citep{kal94,kal95}. Beginning at $\sim$200 AU (10")
projected radius, the NE disk extension is brighter, longer,
and thinner than the SW extension by roughly 20\%. 
Asymmetry is also evident in the disk vertical height;
in the SW the disk height is greater north of the midplane
than to the south, whereas the opposite
is true in the NE extension. The timescale for smoothing
such asymmetries is controlled by the orbital period, and it is
therefore short relative to the age of the star. Hence the observed
asymmetries must be young. 
 
Gravitational perturbation by a brown dwarf companion \citep{whi88} or a close
stellar flyby \citep{kal95} has been suggested as one possible mechanism
for generating large-scale disk asymmetry.
However, no stellar object physically associated with $\beta$ Pic has 
yet been identified.
The key problem in identifying a candidate perturber near $\beta$ Pic
is that the star is bright ($V$=+3.8 mag). Even when  
we artificially eclipse $\beta$ Pic using a coronagraph,
spurious instrumental features that
resemble both stars and extended objects dominate optical
images. However, data obtained at
different telescopes with different instruments have
different instrumental noise signatures. We therefore search for
the faintest real objects in $\beta$ Pic's field 
using optical data from multiple telescopes and instruments.

\section{OBSERVATIONS AND RESULTS}

Table 1 summarizes the observational data
for $\beta$ Pic.
The Hubble Space Telescope (HST) Archive data consist
of WFPC2 CCD images with the $\beta$ Pic
disk oriented along either the Planetary Camera (PC) or the 
Wide Field Cameras (WFC).
The ground-based
observations and data reduction techniques are
described in earlier work \citep{kal95,smi84}.
Here we note that sensitivity to faint objects is hampered by 
$\beta$ Pic's bright point spread function (PSF)
and light from
the circumstellar dust disk itself. Both are
subtracted using template stars and/or idealized model fits that
are detailed in \citet{kal95}.

Subtraction of a smooth, symmetric (axially and vertically)  model disk 
reveals a
brightness enhancement along the NE
extension of the disk midplane 785 AU from the star (Feature A; Fig. 1; Table 2).  
Four field stars common to every data set are utilized for image
registration and astrometry, and the feature is confirmed in every
data set (except the HST PC images, which lack a suitable field of view).  
We do not detect a similar 
brightness enhancement anywhere between 25" and 41" radius on the
SW side of the disk.  Feature A is unlikely
to be a background galaxy because it is amorphous in
the unbinned HST images, and appears extended along
the position angle of the disk.
 
The NE disk midplane has several more brightness
enhancements between 25" and 40" with a degree of positional
correlation not present in brightness knots in the field or SW of the star (Fig. 1; Table 2).
The centroids of Features F and G are spatially correlated to within 0.5" in
the highest resolution images obtained with HST and the University
of Hawaii 2.2 m telescope.  The morphologies and positions of feature centroids vary 
due to instrumental noise, sub-pixel registration differences between   
the observed disk and the idealized model disk, and the
number of pixels binned to form a final image.  
Experiments with the data reduction techniques
show that the latter two effects can shift centroids by
1 pixel, which is equivalent to 0.4"-0.6".  
These uncertainties are evident even for the highest signal-to-noise
detection, Feature A (Fig. 1).  
We identify four more brightness enhancements (B, C, D, E; Fig. 1)
that appear spatially
correlated within these uncertainties.  We expect that more sensitive,
high resolution observations dedicated to imaging this portion of the
disk will help establish the exact number of knots and their positions.
Brightness knots
near and within the SW disk extension are uncorrelated and can be attributed
to noise.

Table 2 gives the measured positions and surface brightnesses of the seven 
features.   Basic characteristics of the midplane structure are:  
a) feature A at $40.7$" (785 AU) radius is resolved and
extended by $\sim$4" ($\sim$80 AU),
b) feature A shows the greatest enhancement over the mean
midplane surface brightness (i.e. displays the greatest contrast),
c) the spacing between features generally increases with increasing radius,  
d) the SW disk midplane does not contain comparable features.
 
We interpret the observed features as dust density enhancements along the
projected disk midplane.\footnote{Given the number of faint (23 mag$< m_R <$24 mag)
objects detected
in the entire unobstructed field of view, there is a $\sim$3\% chance
that one of the disk features is due to a background galaxy.}
The density enhancements may represent discrete clouds of 
dust produced by random planetesimal collisions, or a ring system
system viewed edge-on.
We reason that if random planetesimal collisions were important,
then we would observe density enhancements on the SW side of the disk also.
Because density enhancements are absent in the SW midplane, 
the random collisions mechanism may be insignificant, though future work should
perform quantitative tests.

Here we explore the validity of a ring model 
to explain the midplane density enhancements.
We assume that the multiple features
along the NE midplane represent nested eccentric rings viewed
close to edge-on. 
The absence of comparable features to the SW of $\beta$ Pic means
that the rings are not centered on the star.
The well ordered nature of the system of bright
features over a radial scale $\sim$300 AU may be indicative of
a global restructuring of the planetesimal disc in the recent past.
Thus, as a first attempt to explain the origin of such a system,
we test how a strong gravitational
perturbation, such as from a close stellar flyby,
might alter the global morphology of an initially symmetric 
planetesimal disk. In addition, we require that the
resulting disk morphology qualitatively fits the radial
and vertical disk asymmetries known to exist along the 
same region of the disk as the proposed ring system.

\section{DYNAMICAL SIMULATIONS}

We utilize a standard
numerical code\footnote{
We assume an initial $r^{-3/2}$ radial dependence of surface
number density, and a vertically exponential density profile with the disk
flared such that the scale height is proportional to $r^{3/2}$.}
 \citep{mou97} to follow $\sim 10^6$ collisionless
test particles that
are initialised in circular orbits about a point-mass potential;
this system experiences an encounter with a secondary point-mass
that follows a prescribed parabolic trajectory.
The key parameters governing
the disk response are: the mass ratio, the pericenter
distance $q$, and the inclination of pericenter to the initial
midplane of the disk $i$.
The test particles are taken to represent the underlying parent bodies
that replenish the dust through infrequent
collisional disruption.
We 
assume that the distribution of simulated
particles traces the distribution of dust grains that would result
from collisions in the real system \citep{mou97}.  
 
We find that coplanar encounters distort disk structure near periastron,
leading to the development of transient kinematic spiral features.
These are regions of eccentricity growth. Within a few
disk orbital periods of periastron passage the spiral patterns collapse
into generally eccentric nested rings.  If the encounter timescale
is sufficiently short, as in relatively close encounters,
then one spiral arm dominates the response. 
Correspondingly, the
disk takes on a lopsided appearance \citep{lar97}. 
After many orbital periods, the ring patterns
become
incoherent through orbital phase-mixing.
Inside a radius $\sim q/4$
the tidal influence of the perturber is slight; outside a radius
$\sim q/3$ the interaction rapidly increases in strength \citep{hal96}.
Similarly, non-coplanar flyby encounters excite inclination changes in the
orbits of disk particles \citep{cla93}, generating
asymmetry about the midplane \citep{ost94}.

The stellar mass ratio is taken to be $0.3$ 
throughout the models presented here.
The perturber therefore has
the mass of an M star relative to $\beta$ Pic.  
However, varying the mass ratio was not found to affect the qualitative
outcome significantly compared with varying $q$.
We investigated the induced length asymmetry
in the disk as a function of $q$
for coplanar encounters. For a broad range of viewing angles, the required
length asymmetry ($\sim 20\%$)
is produced at a radius corresponding to the initial size of the disk
when $q$ is $1.3$ times larger than that value.
We then considered various inclinations, $i$, for the
encounter, and compared the resulting isophotal contours with those
from observations, for various viewing angles. 
 
Figures 2, 3, and 4 present the simulation which best reproduces
the observational data.  The model disk is seen $9$ orbital periods 
(at the initial outer radius)
after periastron. 
Planetesimals have scattered outwards to several times the
initial disk radius, and show
well-defined ringed structure on one side of the disrupted disk,
but not the other, in agreement with the observations (Fig. 1).  In the
edge-on view, model disk isophotes qualitatively match the observed 
asymmetries in length, width,
and height above and below the midplane (Fig. 3).  The rings appear as
bumps along the disk midplane in the edge-on view, with spacing
increasing with radius (Fig. 4).   
 
Ring structure inside computational radius 2
has been eliminated by orbital phase 
mixing, which in the observational data corresponds
to radius $\sim$ 26" (500 AU).  
The region of relatively unperturbed particles, $q/4$, now 
scales to $\sim$ 9", which is consistent with
the projected radius  
where disk asymmetries are observed to begin \citep{kal95}.
Having determined the length unit we deduce that this model is at a state
$\sim 90000$ yr after periastron.  Since the removal timescale of dust due to
Poynting-Robertson drag is significantly longer than
10$^5$ yr \citep{bap93}, the distribution of parent bodies
will trace the reflecting dust particles, as we initially assumed. 
 
\section{DISCUSSION}

The stellar flyby hypothesis provides a simple explanation for
the existence, spacing, and morphology of the brightness maxima
along $\beta$ Pic's midplane, as well as
the large-scale disk asymmetries.
The pumped up velocity dispersion of planetesimals may also 
lead to increased dust replenishment rates, owing to more  
frequent and destructive collisions \citep{ste97,ken99}. Therefore, the
large total dust mass around $\beta$ Pic
compared to that of other main sequence stars probably results from
both its youth \citep{byn99} and the dynamical state of the system. 
Our model for $\beta$ Pic's recent dynamical history implies
that the top panel of Fig. 2 approximates the face-on view
of the planetesimal disk.
 
The encounter distance
assumed in the simulation scales to $\sim$700 AU, 
implying a statistically unlikely event 
(0.01\% chance in 10$^6$ yr,
assuming empirical parameters
for the Solar neighborhood; Garcia-Sanchez et al. 1999).
However, if $\beta$ Pic formed with a bound stellar companion
on a $\gg$1000 AU radius orbit, and it was the
companion that was perturbed by a stellar flyby into a new orbit 
that disrupted the planetesimal disk, then the flyby probability
increases by an order of magnitude or more.
This scenario is also consistent with
the prograde and relatively small angle trajectory of the perturber 
relative to the disk midplane in our simulation.
Ultimately, identification of the perturber is necessary to
validate the stellar flyby hypothesis. 
Future work will present a comprehensive, statistical
analysis of the relative space motions
and uncertainties for candidate
perturbers using Hipparcos data \citep{kdl00}.

We note that the orientation of the outer disk
height asymmetry is in the same direction relative to the 
midplane as the proposed midplane warp imaged 
$\sim$50 AU from the star \citep{bur95}. This inner warp could be due to 
a planet that is inclined relative to the disk
midplane \citep{mou97}. 
The stellar flyby hypothesis offers an alternate explanation for
the warp.
The non-coplanar flyby generates
planetesimal orbits with increased eccentricity and
inclination, which at apastron manifest as the flared disk
in the SW extension.  However, as seen in the edge-on
projection, this family of vertically scattered planetesimals 
cross the main disk on their way to
periastron at $\sim$50 AU to the NE of $\beta$ Pic
(Fig. 2).
This intersecting 'second plane' might contaminate
isophotes from the quiescent inner disk, giving the appearance of a
warped midplane.  
Hypothetical planets near 50 AU radius may also encounter
the 'second plane' planetesimals
and deliver them to the innermost parts of the system, leading to an
enhancement of cometary activity \citep{kna94,beu96}.
 
The main effect of the external perturber is to
scatter planetesimals from their formation sites outward to
greater radii, and vertically away from the midplane. 
Roughly 10\% of the disk mass is actually lost to interstellar
space in the simulation presented here.\footnote{
Escaping particles have velocities $\sim$5 km s$^{-1}$, in which case
they could cover the distance between $\beta$ Pic and the Sun
in $\sim$4 Myr.
}
This rearrangement of the planetesimal disk was achieved primarily by the
gas giant planets
during the evolution of the Solar System. However, recent theoretical work
links the high eccentricities and inclinations of Kuiper Belt objects to
close stellar flybys with the young Sun \citep{ida99}.  The current state 
of the $\beta$ Pic system may therefore accurately represent an early evolutionary
phase of our Solar System.



\acknowledgments

P.K. and J.L. acknowledge support from both STScI and MPIA-Heidelberg.
We thank A. Evans, F. Bruhweiler, C. Burrows, S. Ida, J. Surace, R. Terrile 
for contributing to this research.
We are grateful to D. Jewitt, D. Backman, S. Beckwith,
M. Clampin, and J. Papaloizou for commenting on drafts of the manuscript.





\clearpage




\figcaption[fig1]{Two of the four data sets 
used to identify real substructure
along the NE extension
of $\beta$ Pic.  North is up, East is left, and the field of
view for each image is 20"$\times$20".  The left panel shows the
WFC2 data  and the right panel shows the Mauna Kea data (Table 1).
Images are registered using field stars located throughout the
entire fields of view (roughly 200"$\times$200").  
The WFC2 data are binned
to 0.4"/pixel, and smoothed with a Gaussian profile having
$\sigma$ = 0.7 pixel, in order to both match the pixel scale
of the Mauna Kea data, and increase sensitivity to 
nebulosity.
Originally the images contained the NE extension of 
$\beta$ Pic's circumstellar disk. The disk was subtracted
using a model of scattered light from a dusty circumstellar disk fit
to the global morphology of each disk extension.
After subtraction, most of the disk disappears except for radial substructure 
that is not locally fit by the idealized, smooth model disk.
We draw horizontal 
lines indicating midplane features 
that appear spatially correlated to within 2 pixels in all the data listed in Table 1. 
The effects of noise in modulating the position and morphology of features is
evident even for the most robust detection, Feature A.
Features F and G also display 1-2 pixel spatial correlation,
but validating the existence and spacing of
features B, C, D, E requires observations with better
sensitivity and resolution. 
\label{fig1}}

\figcaption[fig2]{
The face-on and edge-on views of the
simulation particles in the computational frame centered on
the $\beta$ Pic mass. The horizontal axes correspond to
the line of nodes of the perturber orbit. Parameters are: $q= 2.6$,
$i= 30\degr$. The disk is initially distributed between $0.2$ and
$2$ radius units. The path of the perturber is indicated with a
dashed line and enters the field from the right hand side.
{\it Top panel}: The face-on view showing the large scale structure of
the perturbed disk. We plot a random sample of a quarter of the particles
outside a radius corresponding to the initial disk extent.
The sense of rotation is counterclockwise, the perturber reaching
closest approach on the vertical axis.
{\it Bottom panel}: The edge-on view produced by rotating
the top panel $90\degr$ about the horizontal axis.   
One side of the disk is vertically distended and strongly
truncated, whereas the opposite side is radially distended
with rings.   
Also, the disk is slightly tilted with respect to its initial configuration
(i.e. midplane along the horizontal axis).
Viewed with the appropriate orientation, we identify the short and
long sides of the disk with $\beta$ Pic's SW and NE extensions, respectively.
\label{fig2}}

\figcaption[fig3]{
{\it Top panel}: Isophotal contours for the simulated disk.
The contours are created by projecting the particle
positions onto a plane perpendicular to the simulated line of sight
and binning particles. The brightness for each bin is the sum of
contributions from each particle in the bin, assuming isotropically
scattered stellar light without extinction. This procedure is used
only as a rough guide to the morphology that would be seen in a
real system in which the scattering surface is provided by dust
supplied from planetesimal collisions.
The appearance of these contours was investigated for various viewing
angles and we considered an azimuthal displacement of $165\degr$
and inclination to line of sight of $4\degr$ to yield the most
satisfactory result. We note that the best viewing angles will
depend on the details of the scattering phase function, the
initial disk model, and the dust evolution model.
{\it Bottom panel}: The observed surface brightness isophotes of the
$\beta$ Pic disk for the 12 Oct. 1993 coronagraphic data. Isophote
spacing is 1 mag arcsec$^{-2}$ and the outer contour gives 
22 mag arcsec$^{-2}$. The NE extension is oriented to the left hand side
of the occulted star, and the north side of the disk midplane is
upwards on the page. 
The bar represents 5" ($\sim$100 AU). The SW side of the disk is
shorter and wider than the NE side, and isophotes extend higher north
of the midplane compared to the south.
\label{fig3}}

\figcaption[fig4]{
{\it Top panel}: A close-up view of particles
in the simulated NE extension.
{\it Bottom panel}: The binned brightness profile along the
simulated NE extension in the top panel. 
The brightness for each bin (0.04$\times$0.08)
was defined by summing 1/r$^2$ for each particle in the bin,
where r is the correct 3-D radius.
The brightness
unit is computational:  brightness from
one particle at unit distance = 1.
If we associate
feature A found in the $\beta$ Pic disk with the
density enhancement at $2.92$ in the simulation, then we can explain the radial
distribution of all bright features, within observational uncertainty.
Arrows indicate real data, appropriately scaled. Error bars
give the $0.6$" uncertainty in the radial position measurements as discussed
in the text.  The density enhancement at radius 3.15 (bottom panel) is
not detected in our data.  It contains roughly half the particles of Feature A
and more sensitive observations are required
to test for its presence.
\label{fig4}}




\clearpage
\begin{table}
\footnotesize
\begin{center}
\caption{Observations \label{tbl-1}}
\begin{tabular}{lcccccr}
\tableline
Telescope& Date & Spot   	& Pixel  & Central & Integration \\
        &       & Diameter   & Scale &     Wavelength & Time   \\
        &      & (arcsec)    & (arcsec/pix)    & (nm) &(seconds)  \\
\hline
Las Campanas (2.5 m) & 02 Feb. 1992& 22.5   & 0.64  & 647 &3900
 \\
Univ. of Hawaii (2.2 m)& 12 Oct. 1993& 6.5    & 0.40  & 647& 655
 \\
HST PC (2.4 m)\tablenotemark{a} & 12 Dec. 1995& none   & 0.05& 674&1200
\\
HST WFC (2.4 m)\tablenotemark{a}  & 15 Oct. 1995& none & 0.10 & 827 & 600 
\\
\tableline
\end{tabular}
\tablenotetext{a}{HST observations
do not use a coronagraph.
The HST PC data have the smallest field of
view such that disk is not detected beyond
32" radius.}
\end{center}
\end{table}

\clearpage
\begin{table}
\begin{center}
\footnotesize
\caption{Northeast Disk Midplane Structure \label{tbl-2}}
\begin{tabular}{ccccc}
\tableline
Feature & Radius\tablenotemark{a} &Surface Brightness\tablenotemark{b} 
& Enhancement\tablenotemark{c}
\\
        & (AU)&  (mag arcsec$^{-2}$) & (\%)                 \\
\tableline
A       & 785  & 23.5 & $>$20   \\
B       & 710  & 23.7 & 10   \\
C       & 647  & 24.1 & 5   \\
D       & 608  & 24.0 & 5   \\
E       & 575  & 24.0 & 5   \\
F       & 543  & 24.0 & 5   \\
G       & 506  & 24.0 & 5   \\
\tableline
\end{tabular}
\tablenotetext{a}{Conversion of radius in arcsec to AU assumes $\beta$ Pic is
19.3 pc from the Sun.  The position of Feature A is the measurement of its
centroid in unbinned HST WFPC2 data (Table 1), with an uncertainty of 5 AU.
The positions of Features B-G are the averages of measurements made in all the 
data listed in Table 1, with uncertainties of $\sim$10 AU.}
\tablenotetext{b}{Absolute and relative surface brightnesses
are sensitive to small changes in the position and scaling of
the axisymmetric model disk used for subtraction.  The estimated uncertainties
in the surface brightness measurements are
$\sigma \sim 0.3$ mag arcsec$^{-2}$ for feature A, and
$\sigma \sim 0.5$ mag arcsec$^{-2}$ for the other features.}
\tablenotetext{c}{Enhancement is the local contrast between a feature and a least-squares,
power-law fit to a 20" segment of the disk midplane.}
\end{center}
\end{table}





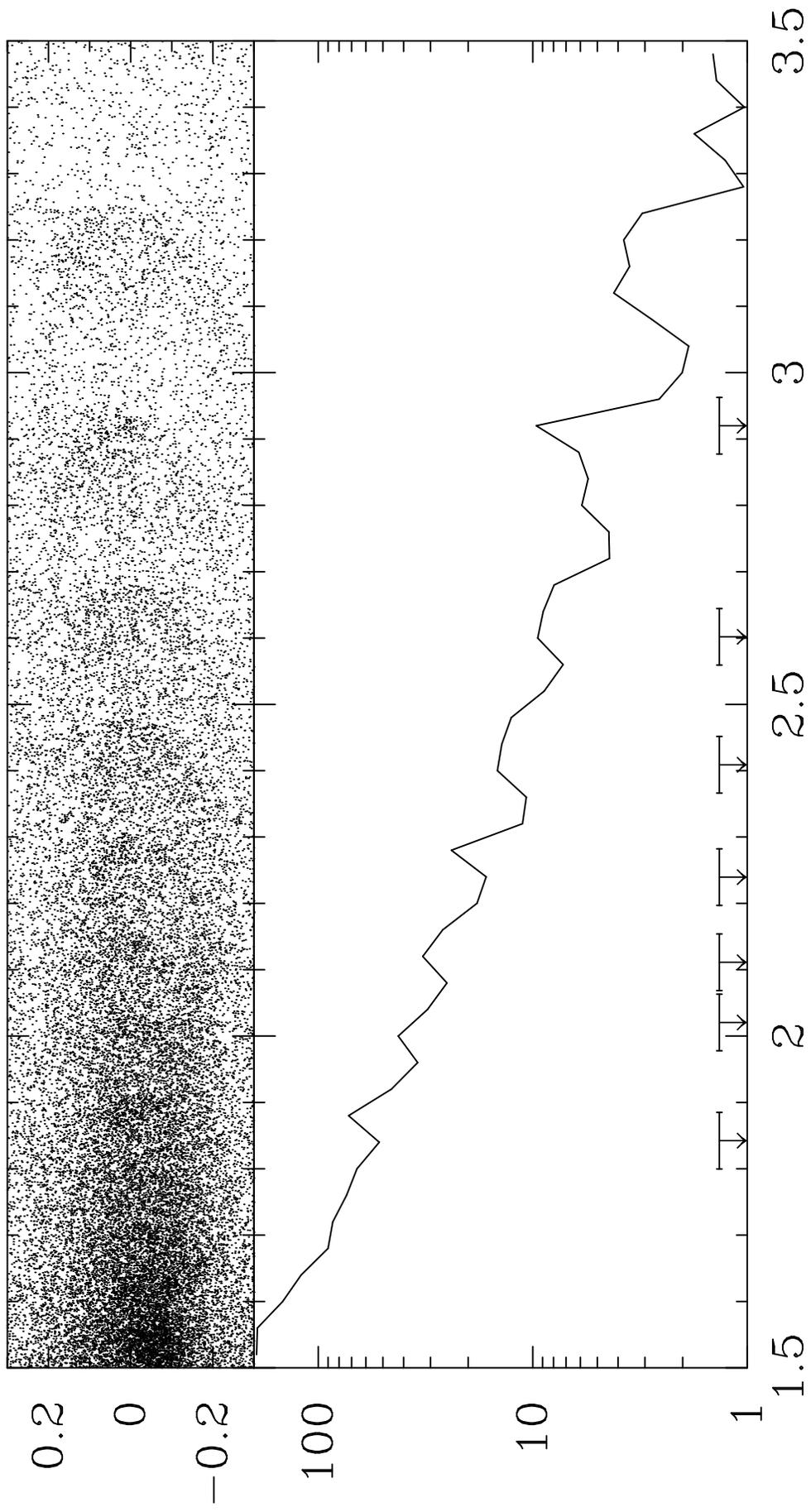
\end{document}